# THE "STRANGE METAL" IS A PROJECTED FERMI LIQUID WITH EDGE SINGULARITIES

ABSTRACT

The puzzling "strange metal" phase of the high Tc cuprate phase diagram reveals itself as closer to a Fermi liquid than previously supposed: it is a consequence of Gutzwiller projection and does not necessarily require exotica such as an RVB or mysterious quantum critical points. There is a Fermi liquid-like excitation spectrum but the excitations are asymmetric between electrons and holes, show anomalous forward scattering and have Z equal to 0. We explain the power law dependence of conductivity on frequency and predict anomalies in the tunneling spectrum, and the forms of photoelectron spectra.

---

INTRODUCTION

Recently we have used the Gutzwiller mean field theory ("Plain Vanilla"[1],[2]) in order to estimate the hole-particle asymmetric tunneling spectra of cuprate superconductors, achieving fairly good qualitative correlation with experiments. When these methods are extrapolated to the case of small or zero superconducting energy gap $\Delta$, one observes that the predicted spectrum acquires a step function singularity at $\omega=0$ (its width is of order $\Delta/(2g)$, g the renormalization factor $\approx 2x/(1+x)$, x the doping percentage). The heights of the tunneling spectra on the two sides, which are also the imaginary parts of Green's functions for local electrons and for local holes, are related as 1(holes):g(electrons).

The only other physical situation in which mean field theory predicts a step-function singularity in the one-electron spectrum is the well-known "x-ray edge" problem of x-rays emitted when an electron from the conduction band falls into an inner shell, and it turned out in that case that in fact the edge singularity was not physical and instead was modified to a power law.[3] (Dispersion theory alone should alert us to this probability since Hilbert transformation of a discontinuity would produce a logarithmic singularity in Re(G), indicating that there are logarithmic divergences in perturbation theory.) The edge singularity in tunneling also cannot stand alone; we suppose that there must be singularities also in the imaginary part of the self-energy, even though the procedures of ref 2 apparently did not predict incoherent parts. Randeria and Zhang[4] deduced from sum rule arguments that the asymmetry must in some sense represent incoherent amplitude, and we confirm that conjecture. We find a close analogy between the two problems in which an edge caused by the Fermi surface appears. The arguments resemble those in[5] but have been tightened and more experimental deductions have been made.

The underlying theoretical concept is quite general for strongly-interacting Fermion systems. When the Mott-Hubbard interaction $U>U_c$, a critical value which for the half-filled single band determines the Mott transition to an insulator, the correct prescription is no longer to renormalize U downward, eventually reaching a Fermi liquid with U=0, but to renormalize U to ∞ by canonical transformation[6], arriving at a state in which double occupancy is projected out, that is, a complete Gutzwiller projection. The canonical transformation leaves us with a Hamiltonian confined to the projected subspace; but we allow that Hamiltonian to act in the full space, looking variationally for the best product solution for the ground state. This Hamiltonian is translation-invariant, and therefore the resulting Hartree-Fock equations determine a spectrum of Bloch wave one-electron states which has a Fermi surface. For well-known reasons, there will be a spectrum of single-particle excitations, taking interactions into account, which will be adiabatically continuable from these, at least close to the Fermi surface.

The true eigenexcitations of the original Hamiltonian after canonical transformation must be confined to the projected subspace. These are not simply the projected quasiparticles, because the projection process leads to strong forward scattering which creates a comoving cloud of soft hole-particle pairs in the original Fermi gas. As a result the projected quasiparticles decay. Whether the decay is into identifiable charge-spin separated eigenexcitations[7] seems unlikely but we leave the question open. Most experimental predictions do not depend on this detail.

An additional difference from conventional Fermi liquid theory comes from the fact that actual physical electrons and holes must be represented as unprojected operators acting on the projected ground state: the physical operators which enter into physical measurements are cP and c*P, and these are not the projected excitations Pc and Pc* we have discussed. In the case of the doped cuprates, commuting c* through P involves a large change in amplitude and is responsible for the asymmetry of the spectrum, while commuting c through P does cause additional incoherence and is partly responsible for the anomalous resistivity and quasiparticle breadth.

FORMALISM

Focussing for definiteness on the hole-doped cuprates, we start from a Hubbard-type Hamiltonian

$$H_{true} = T + U + etc$$
$$= \sum_{in\, j,\sigma} t_{ij} c^*_{i,\sigma} c_{j,\sigma} + U \sum_i n_{i,\uparrow} n_{i,\downarrow} \text{ (+ longer range coulomb, ordinary exchan}$$
[1]

and transform it canonically into the "t-J" Hamiltonian by means of a canonical transformation $e^{iS}$ which eliminates all matrix elements of T which hop an electron onto or off of an already-occupied site. S may be found perturbatively—see ref [5].

$$H_{tJ} = e^{iS} H_{true} e^{-iS} = PTP + J \sum_{inj} S_i \cdot S_j \qquad [2]$$

where P is the Gutzwiller projection operator

$$P = \prod_i (1 - n_{i\uparrow} n_{i\downarrow}) \qquad [3]$$

The Hamiltonian can also be written in terms of the projected field operators

$$\hat{c}^*_{i,\sigma} = (1 - n_{i,-\sigma}) c^*_{i,\sigma}; \hat{c}_{i,\sigma} = c_{i,\sigma}(1 - n_{i,-\sigma}) \qquad [4]$$

and the kinetic energy is then just the sum of products of the "hat" operators.

The terms we have designated as "etc" are also to be understood as acting in the projected subspace only; for our purposes we make the inessential simplification of ignoring them and studying [2] only. Calculations using [2] and renormalized mean field theory quoted in ref [1] show that the superconducting gap vanishes at a doping level of ~30%. We confine ourselves to energies>$\Delta$ or higher dopings than 30%, so that we can neglect the gap; we assume that this is valid everywhere to the right of the T* line in

the conventional phase diagram. The eigenstates of [2] must have the form

$$\Psi = P\Phi(r_1, r_2, \cdots, r_N)$$

and our neglect of $\Delta$, and the assumption that the state is non-magnetic, means that our variational result for the ground state $\Phi_0$ will be simply the product function of Bloch waves

$$\Phi_0 = \prod_{k<k_F} c^*_{k,\uparrow} c^*_{-k,\downarrow} |vac\rangle \qquad [5]$$

(Projection commutes with total number so that $k_F$ will be a conventional Luttinger size Fermi surface.)

The actual manipulations are familiar from reference 1 but our interpretation of them is different from the previous understanding of Gutzwiller theory. We make the point that since $P^2=P$ (for the *full projector P of eq [3] only*), the projected Hamiltonian [2] can be thought of as acting in the full space, on the wave function $\Phi$ and not on its projected part only. Minimization of the energy with respect to all
single-particle excitations leads to the Hartree-Fock equations

$$\langle [H_{tJ}, c^*_{k,\sigma}] \rangle_{ave} \Phi_0 = \varepsilon_k c^*_{k,\sigma} \Phi_0 \qquad [6]$$

(Here we have taken the zero of energy at the Fermi level). We understand the mean value of the commutator to imply that in the definition of the "hat" operators we have replaced $n_\sigma$ by $\langle n_\sigma \rangle$. the hole energies below $k_F$, and the particle energies above, are positive, which is the variational criterion for the Fermi momentum. In the absence of an energy gap

$$\varepsilon_k - E_F \approx v_F(\hat{k}) \cdot [k - k_F(\hat{k})] \quad [7]$$

near the Fermi surface for a general k.

We would emphasize that the t-J Hamiltonian acting in the full space is not at first expected to be very singular. Its kinetic energy matrix elements are relatively sparse, so that the average kinetic energy is renormalized as observed in the RMFT of ref 1. It is reasonable to expect that it has a Fermi level which is sharp, because of the conventional "Landau" arguments from the exclusion principle, and the deviations from the mean can in principle be treated perturbatively. *But the excitations we have derived above are not true quasiparticles in the Landau sense.* The reason is that the t-J Hamiltonian retains in its projective character the residue of the very strong interaction of the Hubbard model.[8]

That is, the operators which create the excitations which have the sharp momentum and Fermi surface are $Pc^*_{k>kF}$ and $Pc_{k<kF}$, while the operators which represent a physical electron or hole tunneling in, or which are excited by an electromagnetic field, are c*P and cP, i e they are real electrons and holes operating on a projected state.

These are simplest in local form. For electrons,

$$c^*_{i,\sigma} P\Phi = (1-P)c_{i,\sigma}{}^* P\Phi + Pc^*_{i,\sigma} P\Phi$$
$$= Pc^*_{i,\sigma} \Phi + \text{high energy parts} \quad [8]$$

and the high energy parts are projected out. But since c* was normalized, any physical probability involving it must be renormalized by the factor x=doping% (see ref [4]) which represents the probability that site i is not already occupied. But once the electron is in, it is in the true single-particle excitation state.

For holes, necessarily the final state will be in the allowed manifold. But it will not be the true excitation.

$$c_{i,\sigma} P\Phi = Pc_{i,\sigma}\Phi - Pc_{i,\sigma} n_{i,-\sigma}\Phi$$
$$= (1 - n_{i,-\sigma})Pc_{i,\sigma}$$

[9]

and the second term in principle can generate three excitations, two holes and an electron. In ref [4] we dealt with the second term by taking its average, and in the mean, therefore,

$$c_{i,\sigma} P \cong Pc_{i,\sigma}(1 - <n_{-\sigma}>)$$

[10]

The ratio of the two amplitudes is $2x/(1+x)=g$, so that we have the dilemma of a step function in the mean field tunneling spectrum for tunneling into a single site at $\omega=0$, as described in the introduction.

It is essential to try to calculate the tunneling spectrum taking the projection process into account. The spectrum of point contact tunneling is the frequency Fourier transform of the imaginary part of the local Green's function for a site 0, $G_{00}(t)$, where

$$G_{00}(t) = \langle 0|c_0(t)c_0*(0)|0\rangle \text{ for electrons}$$
$$= \langle 0|c_0*(t)c_0(0)|0\rangle \text{ for holes, and}$$
$$|0\rangle = P\Phi$$

(11)

Starting with the electron case, we write

$$c_{0\uparrow}*(0) = Pc_{0\uparrow}* + (1-P)c_{0\uparrow}* \cong \hat{c}_{0\uparrow}*(0),$$

recognizing that the second term will necessarily carry us into the forbidden high-energy subspace and must be dropped. The Green's function is thus that of the "hat" operator, an operator which adds one extra projective constraint onto the definition of the ground state: it must have no down-spin electron on site 0.

Calculating this Green's function is surprisingly easy. The "hat" operator is the product of two operators, $c_{0\uparrow}*$, which operates on up spins, and $(1-n_{0\downarrow})$, which operates on down spins only. Our basic approximation is to calculate the propagators using the mean field excitations, that is, the first stage in a successive approximation procedure. But the mean field excitations are a set of independent up- and down-spin quasiparticles, so that the Green's function is simply the product.

$$G_{00}(t) = G_{00,free}(t) \langle 0_\downarrow | e^{-iH_0 t}(1-n_{0,\downarrow})e^{iH_0 t}(1-n_{0,\downarrow}) | 0_\downarrow \rangle,$$

where $H_0$ is the kinetic energy [7] of the mean field down-spin quasipa[rticles]

Let us call the down-spin object multiplying the free quasiparticle $G_{oo}$, $G^*(t)$ It is not hard to show that this is closely related to the "x-ray edge" Green's function introduced by Nozieres and de Domenicis[9] --although the actual form we use was introduced by Doniach and Sunjic[10] and perhaps earlier by Yuval and myself[11]. We can rewrite it as the overlap integral between a state in which the down-spin electron has been excluded from site 0 at time t=0, and then the state propagates according to the mean-field Hamiltonian until time t; with that in which the ground state propagates unperturbed until time t, and then site 0 is excluded:

$$G^*(t) = \langle (1-n_{0,\downarrow})e^{iHt}|0\rangle, e^{iHt}(1-n_{0,\downarrow})|0\rangle\rangle$$

[12]

But the projection operator acting on the ground state can be thought of as an eigenstate of a modified local potential with an infinite repulsive potential at site 0. We instantaneously turn that potential off and allow the original Hamiltonian to operate and the waves to propagate until time t. Then we turn the modified potential back on and calculate the overlap with the projected ground state. This is the inverse of the canonical "x-ray edge" process, and gives exactly the same result. The starting function has a phase shift in each down-spin wave function such that all are perfectly reflected and cannot have any amplitude on site 0; and then we permit the resulting wave functions to propagate until time t, at which we measure the overlap with the original state. As t →∞ G becomes proportional to the ground-state to ground state overlap, which vanishes due to the "orthogonality catastrophe".[12] The physical process is clearest in the representation of Schotte and Schotte[13] who describe it as a shift in the coordinates of the Tomonaga bosons in the symmetric radial channel, which then propagates out from the site at the Fermi velocity.

The Green's function may be calculated by the methods of references 8-10 and 12, and aside from a non-universal constant is well-known to be proportional to

$$G^*(t) \propto t^{-\frac{1}{2}(\delta/\pi)^2}$$ [13]

Here δ is the phase shift at the Fermi surface caused by the effective local potential necessary to empty the site of electrons. This is easily calculated from the Friedel sum rule: since the site contains just ½(1-x) states below the Fermi surface it must be

$$\delta = (1-x)\pi/2 \text{ and the power is } p = 1/8(1-x)^2$$
[14]

[13] must be multiplied by the Green's function for the free up-spin particle, which for a Fermi surface is

$$G_{00,free}(t) \propto 1/t \text{ , so that } G_{00,el}(t) \propto t^{1+p}$$
[15]

(Of course, the divergence at t=0 is not physical, being cut off at a time t of the order of the inverse bandwidth.) To get the tunneling spectrum we Fourier transform, and find

$$G_{00}(\omega) \propto (i\omega)^p \text{ and } \therefore N_{el}(V) = \text{Im} G_{00}(\omega = V) \propto V^p$$
[16]

Thus careful low-temperature measurements on pure samples will display a very sharp zero-bias dip.

The hole side is somewhat more complicated. The relevant Green's function is given in [11], and the hole operator is [9], so that the Green's function we seek is

$$G_{00,holes}(t) = (e^{-iHt}(1-n_{0,\downarrow})Pc_{0,\uparrow} e^{iHt}|P\Phi\rangle, (1-n_{0,\downarrow})Pc_{0,\uparrow}|P\Phi\rangle)$$
[17]

Again, we approximate the excitation spectrum by separate spin up and spin down systems of Fermions, so that the factor $(1-n_{0,\downarrow})$ may be extracted and replaced with G*(t). But then we still have to deal with the Green's function of $Pc_0$. This seems to be identical with that of $Pc_0^*$, in fact the two are simply related by time reversal. Physically, in this case again we cannot insert the hole into a state with a down-spin electron in it, so must empty the 0 site causing a repulsive boundary condition to be suddenly imposed. As far as we can see, then, there are *two* factors of G*(t) and the Green's function [17] becomes

$$G_{oo,holes}(t) \propto t^{-1-2p} \quad [18]$$

Fourier transforming, we obtain

$$G_{oo,hole}(\omega) \propto (i\omega)^{2p} \text{ and } \therefore N_{hole}(V) \propto V^{2p}$$

[19]

We have not computed the amplitude factors. These will presumably have the same ratio between electrons and holes as in mean field theory. Thus we predict an asymmetric tunneling spectrum, with power law anomalies at zero voltage, on both sides.

DISCUSSION

There is a slight inconsistency in our reasoning: We assume a sharp Fermi surface for the propagating quasiparticles in order to derive the power laws for the local Green's functions. To be rigorous, we have to admit that this is a successive approximation procedure and we have definitely not proved convergence.

But there does remain some hope for accuracy. It is possible, indeed likely, that there is a sharp Fermi surface in the sense that the spin bosons in Haldane's tomographic representation may not be affected by the projection and can define a sharp if hidden

Fermi velocity in the spin sector. In what sense this implies the existence of spinons we don't know nor, actually, care.

TRANSPORT CONSEQUENCES
From the local Green's functions [15] and [18] we can deduce the Green's function in momentum space with one additional argument: That the edge singularity effect results from density fluctuations in the form of soft electron-hole pairs propagating in the symmetric ("s") channel emanating from each particular site, as it does in the standard derivations of the x-ray edge problem. The characteristic of such soft excitations is that they all travel with the same velocity when moving in any given direction, namely the Fermi velocity of equation [7]. In the ideal case of an isotropic system, this velocity is constant and one has circular s-waves centered on the original site, but in the more general case one can identify a single channel which is a superposition of waves from the whole Fermi surface. The decay of such circular waves with time as they propagate do to the creation of soft e-h pairs, which we deduced for the origin-to origin component, must be common to the entire circular wavefront, since it represents simply the overlap of the down-spin wave function at time t compared to that at time 0.

The original free-particle origin-to-origin Green's function varies with time as 1/t. It is not hard to see that this variation is the r=0 limit of the general expression,

$$G_{free}(0,r;t) \propto [t - r/v_F(\hat{r})]^{-1} \varphi_0(k_F r - v_F t)$$

[19]

which expresses the fact that electrons traveling in a given direction have only one velocity, so that t can only appear in the expression t-r/$v_F$. ($\varphi_0$ is the appropriate normalized radial wave function for the dimension; it would be exp(ikr) in one dimension.)

At this point we appeal to Huygens' principle: that a plane wave can be thought of as the superposition of circular wavelets emanating from all points on any previous position of the wave-front. (it might be more modern to appeal to Feynman path-integrals.) If the circular waves decay with time as $t^{-p}$, so must the plane waves, and we should still multiply by $G^*(t)$:

$$G_{el}(0,r;t) \propto [t - r/v_F(\hat{r})]^{-1} t^{-p} \varphi_0;$$
$$G_{holes}(0,r;t) \propto [t - r/v_F(\hat{r})]^{-1} t^{-2p} \varphi_0 \quad [20]$$

To my knowledge there are no data on gapless cuprates with which to compare these. It is not very meaningful to try to compare them with nodal measurements on superconductors because the spread of the spectrum in momentum space would inevitably confuse one with neighboring gapped regions. In the future one should introduce the gap as a low-frequency cutoff of the anomalies.

We can use the same physics and the same simple methods we previously used[14] to calculate the conductivity, since it turned out that the result depended only on the scaling properties of G and not on its detailed structure. Conductivity is a two-particle response function and in order to use [21] we must take advantage of the argument of refs 12 and [15] which distinguishes two regimes of transport, which there were unfortunately misnamed "holon drag" and "holon non-drag". In the holon non-drag regime it was assumed that the correct diagram for the conductivity was the simple bubble consisting of non-interacting electron and hole propagators, with no vertex corrections. The condition for this regime is that the excitations into which the electron or hole decays, i e the soft pairs, are scattered before they can recohere into the original quasiparticle. It was argued that the rate of recoherence is of order $T^2/t$ so that at sufficiently low temperatures there is a clear window for this regime. The references did not note this fact, but this is also a good approximation whenever ω>T, since the rate can never be >T. In this regime, although the anomalous forward scattering does not in principle transfer momentum to the lattice it acts as an effective particle decay process because the decohered states are scattered by the normal processes.

In the holon drag regime, which occurs for rapid impurity or umklapp scattering and low values of p, the vertex corrections, which according to Ward identities should cancel the forward scattering, do indeed act prior to particle decay and restore standard Fermi liquid behavior. In my opinion the region to the right of the "dome" and the "T* line" which is thought of as a reversion to Fermi liquid, may be instead a projective NFL in the "drag" regime.

The conductivity in the "non-drag" regime may be calculated by writing out the simple "one-bubble" diagram with the hole and electron lines representing the physical Green's functions [21] inserted into the formula as presented in, for instance, ref 12:

$$\sigma(\omega) = \frac{e^2}{D\pi m^2 \omega} \sum_k \int d\omega' k^2 [f(\omega'-\omega) - f(\omega')] G_{el}(k,\omega') G_{hole}(k,$$

[21]

where f is the Fermi function. It is a simple matter of power counting to realize that the frequency dependence of the conductivity is

$$\sigma(\omega) \propto (i\omega)^{-1+3p} \qquad [22]$$

The coefficient is easily estimated as in reference 13 from the sum rule (the phase is constrained by causality to be as in [23]). This power law dependence has been repeatedly observed in infrared measurements [16],[17],[18], and the estimates of 3p available from refs. 13-16 are about .35±.1, implying p=.12±.03, a bit larger than expected but in the right range. Obviously there is a need to confirm this with measurements of the zero-bias tunneling anomaly.[19] The angle integrated UPS spectrum is also given by [19] and indeed the early results of Petroff[20] exhibit a power-law like slope and curvature at high energies. The high energy parts of ARPES spectra are also known to suggest a universal power law decay such as would result from

FURTHER DISCUSSION

The above is only a model calculation to elucidate the effects of Gutzwiller projection—which itself seems to be generic to systems with strong repulsive interactions and, crucially, a spectrum of fermions which is *bounded above* in energy. Other oxides, organic metals, and particularly the mixed valence lanthanides and actinides are additional compound systems which may exhibit anomalous forward scattering. As pointed out above, normal-seeming Fermi liquid transport can occur in the "drag" regime where single-particle probes would show anomalies. But where anomalous transport is seen is surely the first place to look—as in $CeCu_6Au$ or $YbRh_2Si_2$[21].

A number of explanations for the Strange Metal—or, alternatively, Marginal Fermi Liquid[22]—have appeared. The penumbra of a quantum critical point is often suggested, with little plausibility as to what the order parameter might be or how it might act over such a wide energy range and simultaneously over the whole Fermi surface. The MFL theory as written is, additionally, incorrect in using a logarithmic rather than a power law description, which contradicts the experiments.

Various suggestions as to the spectrum of scatterers causing the phenomena have been put forward, most recently by Norman and Chubukov[23], who correctly point out that neither lattice vibrations nor some fluctuating order parameter is sensible—they propose a continuum of bosons extending at least to .3 ev, and thereby fit the power law. Presumably they have rediscovered the tomographic bosons of the Fermi sea, so it is not clear that the present theory is completely distinct from theirs. I suspect that if they were to calculate Z for their particles, it would come out near to 0.

Finally, what is going on in the cuprates? When the superconducting gap opens, the Fermi edge ceases to be singular, the tomographic bosons become gapped and the low-energy parts of the quasiparticles cease to have the fractional power behavior: the power law decay ceases at long times and becomes truly Z/t-like. This gives us a delightful explanation as to why many of the superconductor's properties may be calculated as though there were real quasiparticles there—there are, below Tc. But above T* and above the gap energy the quasiparticles experience power law decay: essentially, the line of quasiparticle poles turns into a cut in the complex $\omega$ plane. It is hard to imagine another mechanism which can describe this type of experimental behavior in a reasonable way.

ACKNOWLEDGMENTS

I thank Z Schlesinger, T Timusk, and especially N Bontemps and D van der Marel for extensive discussions of their data over the decades; and V Muthukumar, E Abrahams, M. Randeria and M Norman for lengthy discussions of theoretical points.